\documentclass[]{JFM-FLM_Au}

\newcommand\Ray{\mbox{\textit{Ra}}}  % Ra
\newcommand\Nuss{\mbox{\textit{Nu}}}  % Nu
\providecommand\Rey{\mbox{\textit{Re}}}  % Reynolds number
\providecommand\Pran{\mbox{\textit{Pr}}} % Prandtl number
\providecommand\bcdot{\boldsymbol{\cdot}}

\lefttitle{}
\righttitle{Convection in Coldwater}

\title{\color{black}Effects of temperature-dependent material properties in coldwater Rayleigh--Bénard convection}

\author{Gustavo Estay\aff{1}, Daisuke Noto\aff{1,2,3} \and Hugo N. Ulloa\aff{1}}

\affiliation{\aff{1}Department of Earth and Environmental Science, University of Pennsylvania, Philadelphia, PA~19104, USA.
\aff{2}Faculty of Engineering, Hokkaido University, Sapporo, 060-8628, Japan.
\aff{3}Japan Agency for Marine-Earth Science and Technology (JAMSTEC), Yokosuka, 236-0001, Japan.}

\corresau{Gustavo Estay \email{estay@sas.upenn.edu}, Hugo N. Ulloa \email{ulloa@sas.upenn.edu}}

\begin{document}
\maketitle

\begin{abstract}
Water exhibits an anomalous nonlinear density equation of state (EOS) as it approaches freezing, along with an increase in viscosity ($\mu$) and a decrease in thermal conductivity ($k$). Here we ask: how do these temperature-dependent material properties affect thermal convection in coldwater?
We examine these effects within the canonical Rayleigh--Bénard convection (RBC) system, performing direct numerical simulations with coldwater between $0^{\circ}$C and the temperature of maximum density $3.98^{\circ}$C. 
To disentangle the effects of these properties, we consider two main cases: RBC with a quadratic EOS and constant $\mu$ and $k$; and RBC with a quadratic EOS and temperature-dependent $\mu(T)$ and $k(T)$.
We show that the most significant consequence is a shift in the mean fluid temperature relative to `standard RBC'. The nonlinear EOS and the temperature-dependent $\mu$ and $k$ drive this shift in opposite directions, with the former dominating and producing a net decrease in the mean temperature.
This decrease varies with the Rayleigh number $Ra$, reaching up to $0.08^{\circ}$C at $Ra=10^{8}$.
Additionally, the onset of convection is affected, but the influence from the EOS and $\mu(T)$ and $k(T)$ compensate, resulting in a negligible net effect.
Despite these changes, the Nusselt and the Reynolds numbers follow classical scalings in all cases.
Our results establish how the anomalous material properties of coldwater affect local and global features of convection, evidencing the implications of model selection for studies of cryospheric water bodies.
\end{abstract}

\section{Introduction}

Coldwater convection operates under a thermodynamic constraint that is absent from most fluids \citep{debenedetti2003supercooled}: the density-temperature relationship is nonlinear. Additionally, viscosity increases and thermal diffusivity decreases as temperature approaches freezing. These temperature-dependent material properties reshape buoyancy generation and the diffusion of heat and momentum, thereby affecting the fluid physics of ice-bounded aquatic systems. 
{\color{black}Yet much of the theoretical understanding of thermal convection remains rooted in the works of Oberbeck and Boussinesq \citep{oberbeck1879warmeleitung,boussinesq1903theorie}, which rely on three main elements: (i) density variations are neglected everywhere except in the buoyancy force, leading to an incompressible regime; (ii) a linear EOS for density; and (iii) constant values of viscosity, thermal conductivity and heat capacity.
We remark that some studies refer to any deviation from points (i), (ii) or (iii) as a non-Oberbeck-Boussinesq (NOB) model or effect \citep[e.g.,][]{sugiyama2009flow,macek2023assessing,weiss2024rayleigh}. In contrast, a large body of literature refers to point (i) as the Boussinesq (or Oberbeck-Boussinesq) approximation, while allowing departures from (ii) or (iii) \citep[e.g.,][]{veronis1963penetrative,malkovsky2016thermal,couston2021turbulent}.}

{\color{black}The ongoing rapid melting of the cryosphere and the limited understanding of its consequences have triggered a growing number of studies on coldwater systems whose convective dynamics are governed by a nonlinear EOS \citep{du2024physics}. These studies have examined a wide range of processes, including penetrative convection \citep{toppaladoddi2018penetrative,leard2020coupled,pegler2021convective,wang2021universal,ouyang2025touching,olsthoorn2021cooling,grace2023restratification}, radiatively-driven convection \citep{ulloa2019differential,allum2024simulations,2026_estay_pnas_nexus}, geothermally-driven convection \citep{couston2021turbulent,couston2021dynamic}, and ice-melting dynamics \citep{keitzl2016impact, wang2021growth,guo2025effects,angriman2026collective}, in both fresh and salt water environments. However, these studies have not specifically isolated and quantified the effects of the nonlinear EOS on local and global quantities associated with colwater convection. Furthermore, most of the literature neglects temperature-dependent viscosity and thermal conductivity (hereafter diffusion coefficients) in this context, with few exceptions \citep{ishikawa2000numerical,noto_PRF_2026}.}

{\color{black} In the context of Rayleigh--Bénard convection (RBC)}, temperature-dependent diffusion coefficients are {\color{black}known to} produce asymmetric boundary layers and shifts in the bulk mean temperature \citep{ahlers2006non}. Most studies examining {\color{black}these} effects focus on \textcolor{black}{systems} with large temperature contrasts or materials with strong property variations \citep[e.g., planetary mantle-like conditions, solar dynamics, glycerol, gases:][]{pandey2021non,Wan_Wang_2020_JFM,horn2013non, manga1999experimental}, with less emphasis placed on EOS effects. 
Coldwater convection in the cryosphere, however, is different. The accessible temperature range is narrow \citep[e.g.,][]{yang2021new}, and density anomalies generated by a given temperature difference are comparatively small because the thermal expansion coefficient decreases near the temperature of maximum density $T_\textit{md}$, intrinsically weakening buoyancy forcing. Thus, when buoyancy diminishes by the EOS nonlinearity, even modest variations in viscosity and conductivity may exert a disproportionate influence on {\color{black} convective flows} -- an effect not captured by models {\color{black}adopting} {\color{black} constant material properties}. 
\textcolor{black}{Three} questions follow:

\begin{itemize}
    \item[i] {\color{black} How does the nonlinear EOS affect coldwater convection?}
    \item[ii] {\color{black}How do temperature-dependent diffusion coefficients affect coldwater convection, and do their effects couple to those induced by the nonlinear EOS?}
    \item[iii] What local and global quantities are most impacted?
\end{itemize}

Here we address these questions in a canonical configuration that isolates the underlying physics: `coldwater' RBC system. 
We focus our attention on temperatures between the freezing point $T_\textit{fp}$ and $T_\textit{md}$, thereby excluding cabbeling and penetrative-convection regimes. Within this framework, we perform direct numerical simulations to disentangle the effects of the nonlinear EOS and temperature-dependent \textcolor{black}{diffusion coefficients}.

The manuscript is organised as follows. In \S~\ref{sec:framework}, we introduce the theoretical framework, including the models for temperature-dependent material properties and the governing equations. The numerical methodology is described in \S~\ref{sec:methods}. In \S~\ref{sec:results}, we present the results, with emphasis on both local and global properties of cold RBC. Section~\ref{sec:discussion} discusses three main aspects of the dynamics: changes in the mean temperature, scaling relations, and the separate roles of viscosity and thermal conductivity. Finally, \S~\ref{sec:conclusion} summarises our main findings. In particular, we show that symmetry breaking in the temperature field is a robust and measurable feature of coldwater convection, governed primarily by the nonlinear EOS, with temperature-dependent viscosity and thermal conductivity providing smaller but quantifiable corrections.

\section{Theoretical framework}
\label{sec:framework}
We consider a horizontally periodic fluid layer of thickness $h$ -- shallow enough to neglect pressure effects on the EOS -- bounded by two horizontal planes.
Dirichlet boundary conditions are imposed: zero velocity and constant temperatures at the top $(T_t)$ and bottom $(T_b)$, such that $T_\textit{md} \geq T_t > T_b \geq T_\textit{fp}$.
These quantities allow us to consider a \textcolor{black}{reference temperature $T_r = \left(T_t + T_b\right)/2$} and a temperature scale $T_t - T_b$.
To facilitate the discussion, we define the dimensionless temperature $\theta = \left(T - T_r\right)/\left(T_t - T_b\right)$ from the temperature field $T$, leading to the dimensionless counterparts $\theta_t = 1/2$, $\theta_b = -1/2$ and $\theta_r = 0$.

\subsection{Temperature-dependent material properties}
\label{sec:material}
\begin{figure}
\includegraphics{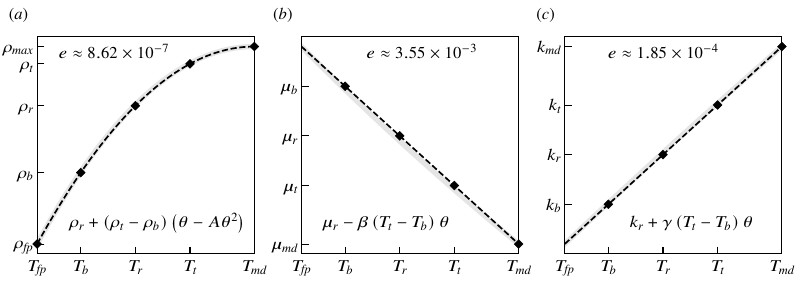}
\caption{Temperature dependence of material properties for cold water, including (\textit{a}) density, (\textit{b}) dynamic viscosity and (\textit{c}) thermal conductivity.
Black dashed lines correspond to the models described in \S~\ref{sec:material}, while key values are highlighted with diamonds, considering example values of $T_b$, $T_r$ and $T_t$.
Actual values of water properties are shown with continuous grey lines  \citep{wagner2002iapws,pythoniapws}.
The maximum relative error of each model is indicated in the corresponding panel, and defined for a general function $f(T)$ as $e = \max_T \left(\left|f - f_\text{model}\right|/\left|f \right|\right)$.}
\label{fig:eos}
\end{figure}

In the following, we describe the models adopted to express the effect of temperature on water density $\left(\rho\right)$, dynamic viscosity $\left(\mu\right)$ and thermal conductivity $\left(k\right)$, which are \textcolor{black}{dimensional quantities} summarised in \hyperref[fig:eos]{figure~\ref{fig:eos}}, {\color{black}using data from the standard IAPWS formulation \citep{wagner2002iapws,pythoniapws}}.
Water characteristics dictate two temperatures of interest, the freezing point temperature $T_\textit{fp} = 0^\circ\mathrm{C}$, and the temperature of maximum density $T_\textit{md}\approx 3.98^\circ\mathrm{C}$.
At these temperatures -- and atmospheric pressure -- freshwater has densities $\rho_\textit{fp}$ and $\rho_\textit{max}$.
In the temperature range of interest, between $T_\textit{fp}$ and $T_\textit{md}$, density can be accurately described by a quadratic EOS 
\begin{equation}
\rho = \rho_\textit{max} - \left( \rho_\textit{max} - \rho_\textit{fp} \right) \frac{\left(T - T_\textit{md}\right)^2}{\left(T_\textit{md} - T_\textit{fp}\right)^2},
\label{eq:eos}
\end{equation}
with a maximum relative error $(e)$ of order $10^{-7}$ as shown in \hyperref[fig:eos]{figure~\ref{fig:eos}}($a$).
{\color{black}We note that this quadratic EOS is equivalent to consider a model with a thermal expansion coefficient that varies linearly with temperature in the range of interest.}
Moreover, in terms of $\theta$, we obtain
\begin{equation}
\rho = \rho_r + \left(\rho_t-\rho_b\right) \left[\theta - A \theta^2\right],
\label{eq:eos_nd}
\end{equation}
where we define the reference density $\rho_r=\rho(T_r)$, and the densities at the boundaries $\rho_t=\rho(T_t)$ and $\rho_b=\rho(T_b)$.
The nonlinear effect in (\ref{eq:eos_nd}) proved to be controlled by the dimensionless parameter
\begin{equation}
A =\frac{2\left[2\rho_r - \left(\rho_t+\rho_b\right)\right]}{\left(\rho_t-\rho_b\right)} = \frac{T_t - T_b}{2\left(T_\textit{md}-T_r\right)}.
\label{eq:A}
\end{equation}

The temperature range allows us to adopt linear models for the viscosity and conductivity:
\begin{align}
\mu &= \mu_r - \beta\left(T_t - T_b\right)\theta,\label{eq:viscosity}\\
k &= k_r + \gamma\left(T_t - T_b\right)\theta,
\label{eq:conductivity}
\end{align}
which are determined by the slopes $\beta$ and $\gamma$.
The reference values $\mu_r$ and $k_r$ are defined as
\begin{align}
\mu_r &= \mu\left(T_r\right) = \mu_\textit{md} + \beta\left(T_\textit{md}-T_r \right), &
k_r &= k\left(T_r\right) = k_\textit{md} - \gamma\left(T_\textit{md} - T_r\right),
\end{align}
where $\mu_\textit{md}$ and $k_\textit{md}$ are, respectively, the viscosity and conductivity at the temperature of maximum density.
For clarity, we introduce the values of viscosity and conductivity at the boundaries: $\mu_b=\mu(T_b)$, $\mu_t=\mu(T_t)$, $k_b=k(T_b)$ and $k_t=k(T_t)$.
We present these models in panels ($b$) and ($c$) of \hyperref[fig:eos]{figure~\ref{fig:eos}}. Note that the maximum relative errors in the linear models, defined as $e=\max_{T}\left(|f-f_{\rm model}|/|f|\right)$ for a function $f$, are of order $10^{-3}$ and $10^{-4}$ for the viscosity and conductivity, respectively.

In addition to the aforementioned material properties, heat capacity at constant volume $(c_v)$ is required to fully describe heat transfer in the system.
This thermodynamic coefficient is also temperature dependent.
However, we neglect this effect and consider $c_v$ constant for the following reasons: (i) The maximum relative variation of this parameter in the temperature range of interest is two orders of magnitude lower than the case of viscosity, and one order of magnitude lower than the case of conductivity;
(ii) The value of $c_v$ does not influence the steady state energetics of the system; it does not affect its steady conductive state either;
(iii) Spatial derivatives of $c_v$ do not appear in the governing equations, in contrast with the case of $\mu$ and $k$, whose spatial gradients have a direct effect on the momentum and heat fluxes.

A dimensionless number $B$ can be obtained directly from equation (\ref{eq:viscosity}), comparing the viscosity contrast with the reference viscosity.
{\color{black}With the aid of $c_v$}, reference values of viscosity and conductivity can be compared through the Prandtl number $\left(\Pran\right)$.
Additionally, we define $C$ to quantify the contrast of conductivity. In sum,
\begin{align}
B &= \frac{\beta\left(T_t - T_b\right)}{\mu_r}
= \frac{\mu_b - \mu_t}{\mu_r}, &
\Pran &= \frac{\mu_r c_v}{k_r}, &
C &= \frac{\gamma\left(T_t - 
T_b\right)}{c_v\mu_r} = \frac{k_t-k_b}{c_v\mu_r}.
\end{align}

\subsection{Governing equations}
\label{sec:equations}
Using $h$ as a length scale, we define a dimensionless Cartesian coordinate system with position vector $\boldsymbol{x} = x\boldsymbol{i}+y\boldsymbol{j}+z\boldsymbol{k}$, so that $z=0$ and $z=1$ correspond to the bottom and top boundaries, respectively.
Then, gravitational acceleration can be written as $\boldsymbol{g}=-g\boldsymbol{k}$, where $g=\left|\boldsymbol{g}\right|$.
From the free-fall velocity scale
\begin{equation}
U=\sqrt{gh\left(\rho_t-\rho_b\right)/\rho_r}
\end{equation}
and its corresponding time scale $\tau=h/U$, we define the dimensionless velocity $\boldsymbol{u}$ and dimensionless time $t$.

The governing equations correspond to the conservation of mass, momentum and energy. {\color{black}Here, we adopt} the main idea behind the Oberbeck--Boussinesq approximation: the temperature-dependent density (\ref{eq:eos_nd}) is used in the buoyancy term of the momentum equation, while the reference density $\left(\rho_r\right)$ is used elsewhere. \textcolor{black}{This approximation is well justified for the temperature range examined in this study (figure~\ref{fig:eos}) and the maximum density variation $\Delta\rho=\rho_\textit{max}-\rho_\textit{fp}$, which, relative to $\rho_{r}$, is about $\Delta\rho/\rho_{r}\approx1.3\times10^{-4}$}.
{\color{black}Thus, the equation of mass conservation in non-dimensional form,
\begin{equation*}
0 = \frac{\partial}{\partial t}\left(\frac{\rho}{\rho_r}\right) +
\nabla\bcdot \left(\frac{\rho\boldsymbol{u}}{\rho_r}\right)
= \frac{\partial}{\partial t}\left(\frac{\rho}{\rho_r}\right) +
\nabla\left(\frac{\rho}{\rho_r}\right) \bcdot \boldsymbol{u} +
\frac{\rho}{\rho_r}\nabla\bcdot\boldsymbol{u},
\end{equation*}
is simplified to the incompressibility condition
\begin{equation*}
\nabla\bcdot\boldsymbol{u}=\boldsymbol{0}.
\end{equation*}
In this framework, the non-dimensional momentum and heat equations are}
\begin{equation}
\frac{\partial \boldsymbol{u}}{\partial t} +
\left(\nabla \boldsymbol{u}\right)\boldsymbol{u}
= -\nabla p_m + \left(A\theta^2 - \theta \right)\boldsymbol{k}
+ \sqrt{\frac{\Pran}{\Ray}} \nabla^2\boldsymbol{u}
- B\sqrt{\frac{\Pran}{\Ray}}
\left[\theta\nabla^2\boldsymbol{u} + {\color{black}\left(\nabla\boldsymbol{u} + \left(\nabla\boldsymbol{u}\right)^\top\right)}\nabla\theta\right],
\label{eq:momentum}
\end{equation}
\begin{equation}
\frac{\partial \theta}{\partial t} + \nabla\theta \bcdot \boldsymbol{u}
= \frac{1}{\sqrt{\Pran\Ray}} \nabla^2\theta
+ C\sqrt{\frac{\Pran}{\Ray}}\left(\theta\nabla^2\theta + \nabla\theta \bcdot \nabla\theta\right),
\label{eq:heat}
\end{equation}
where $p_m$ is a dimensionless modified pressure. The boundary conditions are
\begin{align}
\theta\left(z=0\right) &= -1/2, &
\theta\left(z=1\right) &= 1/2, &
\boldsymbol{u}\left(z=0\right) &= \boldsymbol{u}\left(z=1\right)=\boldsymbol{0}.
\label{eq:bc}
\end{align}

In addition to the dimensionless parameters defined in \S~\ref{sec:material}, the description of the system is completed with the Rayleigh number $\left(\Ray\right)$,
\begin{equation}
\Ray = \frac{gh^3\rho_r\left(\rho_t - \rho_b\right)c_v}{\mu_r k_r}.
\end{equation}

For this work, we focus on the case $T_b = T_\textit{fp}$ and $T_t = T_\textit{md}$.
This conditions correspond to the values $A = 1$, $\Pran \approx 12.626$, $B \approx 0.133$, and $C \approx 0.00138$.

\subsection{Diagnostic variables}
The dynamic response of the system can be quantified by an \textit{a posteriori} Reynolds number, defined as
\begin{equation}
\Rey = \frac{\rho_r h}{\mu_r}
\sqrt{U^2\overline{\left<\boldsymbol{u}\bcdot\boldsymbol{u}\right>}}
= \sqrt{\frac{\Ray}{\Pran}}\sqrt{\overline{\left<\boldsymbol{u}\bcdot\boldsymbol{u}\right>}},
\label{eq:Re}
\end{equation}
where the brackets $\left<\,\cdot\,\right>$ denote spatial average over the domain, and the bar over a function $f$, $\overline{f}$, denotes time average.

The thermal response of the system can be described in terms of the Nusselt number, i.e., the ratio of the actual heat flux to the conductive heat flux at a boundary.
In our case, the conductive state of the system is not described by a linear temperature profile, due to the temperature-dependent thermal conductivity.
Instead, a differential equation for the dimensionless conductive temperature $\varphi$ can be obtained from the heat equation (\ref{eq:heat}),
\begin{equation}
0 = \frac{\mathrm{d}^2\varphi}{\mathrm{d}z^2}
+ C\Pran \frac{\mathrm{d}}{\mathrm{d}z}
\left(\varphi \frac{\mathrm{d}\varphi}{\mathrm{d}z}\right),
\end{equation}
which{\color{black}, considering boundary conditions (\ref{eq:bc}),} can be readily integrated to obtain
\begin{align}
\frac{\mathrm{d}\varphi}{\mathrm{d}z} &= \frac{1}{1 + C\Pran\varphi}, &
\varphi + \frac{C\Pran}{2} \varphi^2 &= z + \frac{C\Pran}{8} - \frac{1}{2}.
\label{eq:conductive}
\end{align}
Then, the Nusselt numbers at the top and bottom boundaries, respectively $\Nuss_t$ and $\Nuss_b$, are
\begin{align}
\Nuss_t &=
\left.
\left<\left.\frac{\partial\theta}{\partial z}\right|_{z=1}\right>_{\color{black}H}
\middle / 
\left.\frac{\mathrm{d}\varphi}{\mathrm{d} z}\right|_{z=1}
\right.
= \left(1 + \frac{C\Pran}{2} \right)
\left<\left.\frac{\partial\theta}{\partial z}\right|_{z=1}\right>_{\color{black}H},\\
\Nuss_b &=
\left.
\left<\left.\frac{\partial\theta}{\partial z}\right|_{z=0}\right>_{\color{black}H}
\middle / 
\left.\frac{\mathrm{d}\varphi}{\mathrm{d} z}\right|_{z=0}
\right.
= \left(1 - \frac{C\Pran}{2} \right)
\left<\left.\frac{\partial\theta}{\partial z}\right|_{z=0}\right>_{\color{black}H},
\end{align}
where the brackets $\left<\, \cdot\, \right>_{\color{black}H}$ denote {\color{black} horizontal average}.
Taking a spatial average and then a time average ($\overline{{\,\cdot\,}}$) of the heat equation (\ref{eq:heat}), we obtain
\begin{equation}
\overline{\Nuss_t} = \overline{\Nuss_b} = \Nuss
\label{eq:nuss}
\end{equation}
allowing us to define the time-averaged Nusselt $\Nuss$.
{\color{black}In practice, it can also be computed as
\begin{equation}
\Nuss
= \frac{1}{2}\left(\overline{\Nuss_t} + \overline{\Nuss_b} \right).
\label{eq:nuss_alt}
\end{equation}
}

Another aspect of the conductive state is that, when $C > 0$, its average dimensionless temperature is nonzero.
In fact, from (\ref{eq:conductive}) we obtain the profile
\begin{equation}
\varphi\left(z\right) = \frac{1}{C\Pran}
\left[-1 + \sqrt{\left(-1 + C\Pran/2\right)^2 + 2C\Pran \, z}\right].
\label{eq:cond_profile}
\end{equation}
As a result, the conductive mean temperature is given by
\begin{equation}
\left<\varphi\right> = \frac{1}{24C^2\Pran^2}
\left[\left(C\Pran + 2\right)^3 - \left|C\Pran - 2\right|^3 - 24C\Pran\right],
\end{equation}
which for $C\Pran < 2$ simplifies to
\begin{equation}
\left<\varphi\right> = C\Pran / 12.
\end{equation}
The result is consistent with the fact that for $C=0$, $\left<\varphi\right> = 0$, which is readily obtained from (\ref{eq:conductive}).
In a convective regime, the effects of the quadratic EOS and the temperature-dependent viscosity also affect the symmetry of the temperature field.
As a result, the average dimensionless temperature of the system $\theta_m = \overline{\left<\theta\right>}$ is a quantity of interest -- it cannot, in general, be expected to be zero as in the {\color{black}standard RBC} case.

\section{Methods and simulations}
\label{sec:methods}

To investigate this system, we perform two-dimensional (2D) direct numerical simulations (DNS) using the governing equations and boundary conditions described in \S~\ref{sec:framework}. While 2D RBC cannot reproduce all aspects of three-dimensional (3D) convection, it retains the key processes relevant to the present study, including buoyancy-driven instability, thermal and viscous boundary-layer coupling, plume generation, and the associated global heat and momentum transport. At moderate and large $Pr$, previous studies have shown that 2D and 3D RBC exhibit broadly similar $Nu$–$Ra$ behaviour, differing mainly in prefactors rather than scaling trends \citep{van2013comparison}. This approximation is particularly appropriate at large $Pr$, where 3D flows become increasingly quasi-2D \citep{pandey2016similarities}. Accordingly, 2D DNS offer an efficient framework for isolating the effects of the nonlinear EOS and variable \textcolor{black}{diffusion coefficients} across a broad range of $Ra$, while preserving the leading-order transport physics \citep{van2013comparison,pandey2016similarities,zhu2018transition}. Similar reasoning has motivated the use of 2D simulations in other thermally driven convective flows requiring dense parameter exploration and large computational domains \citep{ulloa2019differential,ulloa2022development,allum2025radiatively,noto_PRF_2026}.

The governing equations are integrated with the aid of the Dedalus spectral solver \citep{burns2020} in a rectangular domain with an aspect ratio of $4:1$.
We use a Fourier expansion along the $x$-axis, so the system is periodic in the horizontal; whereas on the vertical axis $z$, we use a Chebyshev expansion, so that Dirichlet boundary conditions can be imposed.
The number of \textcolor{black}{horizontal and vertical grid points, $n_x$ and $n_z$,} is selected for each simulation such that the boundary layers and the Batchelor length scale -- the controlling microscale for $\Pran > 1$ -- are properly resolved.

Here, we perform two main sets of numerical simulations.
The first one corresponds to the most general case, namely, variable \textcolor{black}{diffusion coefficients (VDC)} -- $k$ and $\mu$ -- along the nonlinear EOS {\color{black}($A=1$, $B\neq0$, $C\neq0$)}.
The second group keeps the nonlinear EOS, but considers constant \textcolor{black}{diffusion coefficients (CDC)}, that is $A=1$, $B=0$, $C=0$.
Both groups consider the same temperature boundary conditions, having the same values for the parameters $A$ and $\Pran$. 
In both cases, the range of Rayleigh numbers from $10^3$ to $10^8$ is explored.
{\color{black}We also consider the cases ($A=1$, $B = 0$, $C\neq 0$) and ($A=1$, $B\neq0$, $C=0$), to decouple the effects of $\mu$ and $k$.
Additionally, we consider $A=B=C=0$, the standard Rayleigh--Bénard case (SRB) with a linear equation of state and constant diffusion coefficients, for comparison.
A summary of the simulations performed for this work is presented on \hyperref[tab:simulations]{table~\ref{tab:simulations}}, indicating the number of vertical and horizontal nodes.}

{\color{black} The simulation time is defined to allow a statistically robust computation of global quantities at steady-state conditions. Equation (\ref{eq:nuss}) allows us to define the following metric to check that steady-state conditions actually are achieved: $(\overline{\Nuss}_t - \overline{\Nuss}_b)/\overline{\Nuss}_t$, which should be as close to zero as possible.
In our simulations, the value for the metric ranges between $-0.04\%$ and $0.04\%$ as shown in \hyperref[fig:steady]{figure~\ref{fig:steady}}$(a)$. The computation of the Nusselt number using equation (\ref{eq:nuss_alt}) for a representative case is illustrated in \hyperref[fig:steady]{figure~\ref{fig:steady}}$(b)$.}

To estimate the critical Rayleigh number $\Ray_c$ for the aforementioned cases, we perform simulations with decreasing values $\Ray$, progressively approaching the onset.
When the simulations are close enough to the onset, the behaviour of the Nusselt number $\Nuss$ is linearly extrapolated from the two cases with lower $\Ray$, and $\Ray_c$ is found at the theoretical condition $\Nuss = 1$.
For validation, a case with $\Ray < \Ray_c$ is performed to check whether the instabilities originating from a random initial condition decay, and a conductive steady state with $\Nuss = 1$ is obtained.

{\color{black}To estimate the thickness of the thermal boundary layers, we consider a slope-based method \citep{ahlers2006non}. 
That is, the tangent of the temperature profile at the boundary is intersected with $\left<\theta\right>$.
The distance between this intersection and the boundary defines the boundary layer thickness.
Then the thickness of the top $\delta_t$ and bottom $\delta_b$ boundary layers is given by
\begin{align}
\delta_t &= \frac{1/2 - \left<\theta\right>}{\left<\left.\frac{\partial\theta}{\partial z}\right|_{z=1}\right>_{\color{black}H}} ,
& \delta_b &= \frac{1/2 + \left<\theta\right> }{\left<\left.\frac{\partial\theta}{\partial z}\right|_{z=0}\right>_{\color{black}H}}.
\label{eq:bl}
\end{align}
In contrast with equation (\ref{eq:nuss}), $\overline{\delta}_t$ and ${\overline{\delta}_b}$ are not necessarily equal.}

\begin{figure}[t]
\includegraphics{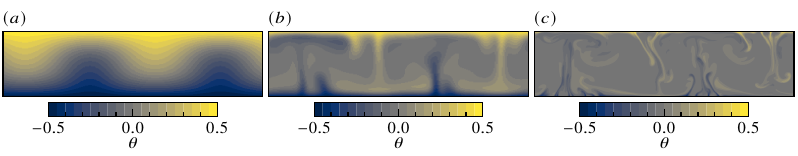}
\caption{Instantaneous snapshots of the temperature field $\theta$ for cases with variable \textcolor{black}{diffusion coefficients} for different values of the Rayleigh number: $(a)$ $\Ray = 2.15\times10^3$,  $(b)$ $\Ray = 2.15\times10^5$, $(c)$ $\Ray = 4.64\times10^7$.}
\label{fig:theta_2D}
\end{figure}

\section{Results}\label{sec:results}
We illustrate different convective regimes in \hyperref[fig:theta_2D]{figure~\ref{fig:theta_2D}} through instantaneous snapshots of the temperature field $\theta$ for cases with variable \textcolor{black}{diffusion coefficients}. For $\Ray = 2.15\times10^3$, the flow exhibits steady convection with prominent diffusive effects, as shown in \hyperref[fig:theta_2D]{figure~\ref{fig:theta_2D}}$(a)$. Increasing the thermal forcing to $\Ray = 2.15\times10^5$ yields unsteady, chaotic convection, as shown in \hyperref[fig:theta_2D]{figure~\ref{fig:theta_2D}}$(b)$, while at $\Ray = 4.64\times10^7$ the convective flow is turbulent, as shown in \hyperref[fig:theta_2D]{figure~\ref{fig:theta_2D}}$(c)$.
{\color{black}We note that, while the results from \hyperref[fig:theta_2D]{figure~\ref{fig:theta_2D}} have the same domain size and aspect ratio in non-dimensional terms, the situation is different for the dimensional counterpart, as the increase in $\Ray$ can be understood as an increase in the layer thickness $h$.}

\begin{figure}[t]
\includegraphics{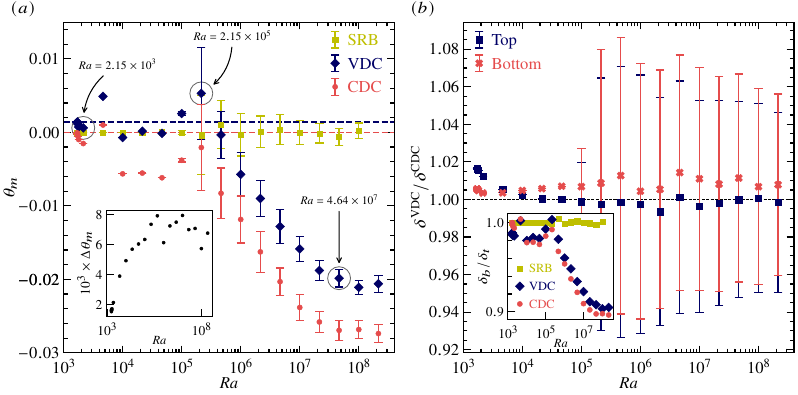}
\caption{$(a)$ Mean temperature $\theta_m$ as a function of the Rayleigh number $\Ray$ for simulations with CDC (red circles) and VDC (blue diamonds) {\color{black}and RBC (yellow squares)}. Vertical bars denote temporal variability through standard deviation at steady state. The horizontal dashed lines indicate the mean temperature for the conductive state for cases with CDC (red) and VDC (blue). Cases that are also shown in \hyperref[fig:theta_2D]
{figure~\ref{fig:theta_2D}} are highlighted with a grey circle.
{\color{black}The inset shows $\Delta\theta_m$, the difference between the mean temperature for the VDC and CDC cases.}
$(b)$ Ratio between boundary layer thickness of VDC and CDC cases $\delta^{\rm VDC}/\delta^{\rm CDC}$ as a function of the Rayleigh number $\Ray$. Blue squares and red crosses are used for the top and bottom boundary layers, respectively. Vertical bars denote temporal variability through standard deviation at steady state.
{\color{black}The inset shows the ratio between bottom and top boundary layer thickness $\delta_b / \delta_t$ following the notation from panel $(a)$ CDC and VDC cases, without indicating temporal variability.}}
\label{fig:mean_theta}
\end{figure}

The asymmetry of the temperature distribution due to \textcolor{black}{the nonlinear EOS and variable diffusion coefficients can be quantified by the mean temperature $\theta_m$, which is zero in a symmetric profile.}
\hyperref[fig:theta_2D]{Figure~\ref{fig:mean_theta}}$(a)$ shows $\theta_m$ as a function of the Rayleigh numbers, comparing cases with constant and variable \textcolor{black}{diffusion coefficients}.
{\color{black}Vertical bars indicate temporal variability, where the highest occur around $Ra=2.15\times10^5$. This high variability is associated with the transition from a steady to a chaotic regime \citep{krishnamurti1973some,hartlep2005transition}.}
For CDC, plotted with red circles, the nonlinear EOS breaks the OB symmetry in a `Rayleigh number dependent' manner: $\theta_m$ is negative over most of the range, except for a narrow interval of positive values at $\Ray = 4.64\times10^3$. 
Overall, the dependence of $\theta_m$ on $\Ray$ is non-monotonic. However, for $\Ray \geq 2.15\times10^5$, $\theta_m$ decreases monotonically, reaching a plateau at $\Ray = 10^8$ with $\theta_m \approx -0.027$.
For VDC, plotted with blue diamonds, $\theta_m$ is systematically higher than in the CDC cases while following a similar trend.
{\color{black}For instance, we have $\theta_m \approx -0.021$ for $\Ray = 10^8$, which correspond to the shift $\left<T\right> - T_r \approx -0.08^{\circ}\mathrm{C}$ in dimensional terms.}
Consequently, the region with $\theta_m>0$ extends to lower Rayleigh numbers, and an additional interval with $\theta_m>0$ appears near $\Ray = 10^5$.
Nonetheless, the mean temperature is negative ($\theta_m < 0$) for most of the studied range, indicating the predominant effect of the nonlinear EOS on the temperature asymmetry -- especially at high $\Ray$.
{\color{black}The latter is emphasised in the inset of \hyperref[fig:theta_2D]{figure~\ref{fig:mean_theta}}$(a)$, which shows the mean temperature difference $\Delta\theta_m$ between VDC and CDC cases as a function of the Rayleigh number.
This difference increases monotonically until $\Ray\sim10^5$, then the trend is less clear, but oscillates around the value $7\times10^{-3}$.}

We further quantify the asymmetry of the temperature profile through the ratio of the bottom to top boundary layer thicknesses $\delta_b / \delta_t$, defined in (\ref{eq:bl}), as a function of $\Ray$.
This ratio exhibits a trend similar to that of $\theta_m$, primarily controlled by the nonlinear EOS -- as shown in the inset of \hyperref[fig:mean_theta]{figure~\ref{fig:mean_theta}}$(b)$.
The largest departure from symmetry occurs, again, at the higher Rayleigh numbers, where $\delta_b / \delta_t\approx 0.9$, corresponding to a $10\%$ difference between the bottom and top boundary layers.
The influence of VDC is also apparent, as observed by the systematic offset of the blue diamonds relative to the red circles.
However, $\delta_b / \delta_t$ alone cannot disentangle how VDC affects the individuality of each boundary layer.
We overcome this limitation by examining the effect of VDC on each boundary layer thickness $(\delta^{\text{VDC}})$ in relation to its counterpart determined from CDC $(\delta^{\text{CDC}})$. 
\hyperref[fig:mean_theta]{Figure~\ref{fig:mean_theta}}$(b)$ illustrates the ratio $\delta^{\text{VDC}}/\delta^{\text{CDC}}$ for the bottom and top boundary layers.
{\color{black} The main finding is that VDC do not affect the top and bottom boundary layers in the same way. While the bottom boundary layer is always thicker, the top one can be either thicker or thinner depending on $\Ray$. We also emphasize that the effects of VDC are secondary when compared with the effects of the nonlinear EOS, with $99.34\%\leq\delta^{\text{VDC}}/\delta^{\text{CDC}}\leq101.66\%$ for the top and $100.34\%\leq\delta^{\text{VDC}}/\delta^{\text{CDC}}\leq101.43\%$ for the bottom boundary layers.}

\begin{figure}[t]
\includegraphics{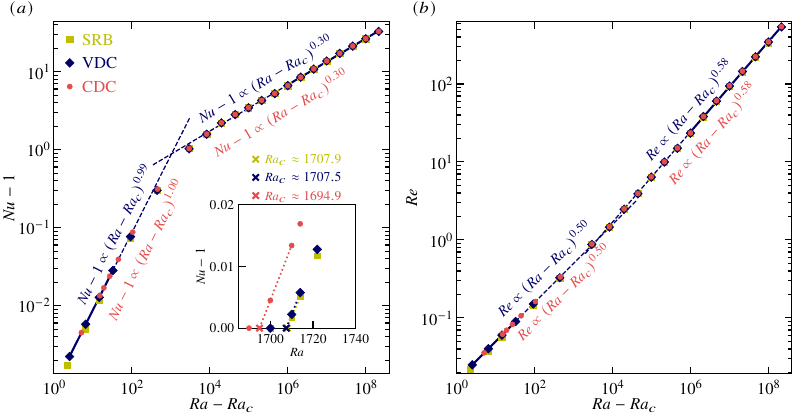}
\caption{$(a)$ Heat transfer rate represented by `$\Nuss -1$' as a function of `$\Ray - \Ray_c$' for \textcolor{black}{SRB (yellow squares),} VDC (blue diamonds) and CDC (red circles) cases. Solid blue lines represent power-law fits for the VPM case across low and high $\Ray$ regions, whereas the dashed lines correspond to their extrapolation.
In the fully steady convective regime, the fit obtained is $\Rey \propto (\Ray-\Ray_c)$; in the turbulent convective regime, the fitting gives $\Rey \propto (\Ray-\Ray_c)^{0.3}$. The inset shows `$\Nuss -1$' as a function of $\Ray$ in a range close to the onset, where the dotted lines indicate the extrapolation used to estimate the critical Rayleigh numbers $\Ray_c$ (see \S~\ref{sec:methods}), which are indicated by crosses on the horizontal axis. $(b)$ Reynolds number, $\Rey$, computed \textit{a posteriori} using (\ref{eq:Re}), as a function of $\Ray-\Ray_c$ for the \textcolor{black}{SRB (yellow squares),} VDC (blue diamonds) and CDC (red circles) cases. Solid blue lines denote power-law fits for the VDC case across low and high $\Ray$ regions, whereas dashed lines show their extrapolation. In the fully steady convective regime, the fit obtained is $\Rey \propto (\Ray-\Ray_c)^{1/2}$; in the turbulent convective regime, the fitting gives $\Rey \propto (\Ray-\Ray_c)^{0.58}$. The extrapolated fits highlight the break in slope around $\Ray-\Ray_c \sim 10^{4}$. }
\label{fig:nu}
\end{figure}

Having established that the nonlinear EOS and the VDC affect the symmetry of the system -- as evidenced by both the mean temperature and thermal boundary layers -- we now turn our attention to energetic quantities, focusing on the heat-transfer rate and the mean kinetic energy as a function of the Rayleigh number. 
We first note that the onset of convection occurs at different critical Rayleigh numbers, $\Ray_c$, in the CDC and VDC cases.
A meaningful comparison, therefore, requires accounting for these shifts in onset.
Accordingly, we analyse the heat transfer, quantified by the Nusselt number $\Nuss$, relative to pure conduction, $\Nuss-1$, as a function of the super-criticality $\Ray-\Ray_c$, as shown in \hyperref[fig:nu]{figure~\ref{fig:nu}}$(a)$.
The results show two distinctive regimes, each well described by a power law: a near-onset regime for $\Ray - \Ray_c < 10^2$ and a strongly supercritical regime for $\Ray - \Ray_c > 4.6\times10^5$.
Power-law fits yield exponents of approximately `1' in the former and `0.3' in the latter, with essentially the same values for CDC and VDC{\color{black}, and no appreciable differences with respect to the SRB case}.
With that said, VDC does have the effect of increasing the critical Rayleigh number to 1707.4, compared with 1694.9 for CDC.
These values are obtained by extrapolating results in the convective regime close to onset, as described in \S~\ref{sec:methods} and shown in the inset of \hyperref[fig:nu]{figure~\ref{fig:nu}}$(a)$.

\hyperref[fig:nu]{Figure~\ref{fig:nu}}$(b)$ shows the Reynolds number, defined in (\ref{eq:Re}), as a function of $\Ray-\Ray_c$.
Consistent with the previous analysis of the heat transfer rate, the behaviour of the $\Rey$ can be effectively described by power laws in the same two regimes.
The region close to the onset exhibits an exponent of 0.5, while an exponent of 0.58 is observed for the strongly supercritical regime.
{\color{black}The latter exponent is not expected for a 3D domain in our parameter space, but it is a known feature of 2D convection \citep{van2013comparison}.}
Again, there is no significant difference in the exponents obtained from the simulations with CDC and VDC.

\section{Discussion}\label{sec:discussion}

\subsection{Effect of nonlinear EOS on mean temperature}

\begin{figure}[t]
\includegraphics{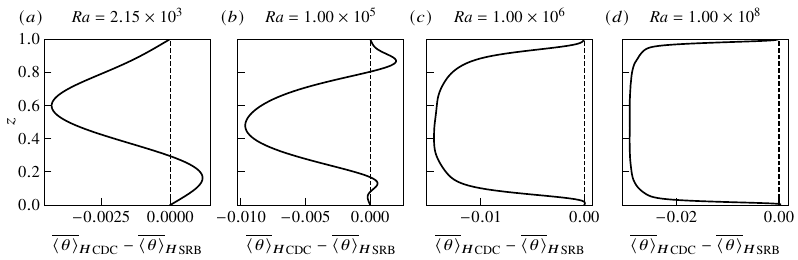}
\caption{Difference between time- and horizontally-averaged temperature profiles for CDC and SRB cases, ${\overline{\langle\theta\rangle}_{H}}_{\text{CDC}}- {\overline{\langle\theta\rangle}_{H}}_{\text{SRB}}$. Panels show increasing Rayleigh numbers, from $\Ray=2.15\times 10^{3}$ to $\Ray=1.00\times 10^{8}$. Vertical dash line denotes zero.}
\label{fig:comparison_rbc}
\end{figure}

In the standard RBC model, which adopts a linear EOS and constant diffusion coefficients, the time- and horizontally-averaged temperature distribution exhibits the familiar `S-shaped' vertical profile, satisfying
$$\overline{\langle \theta \rangle}_{\color{black}H}(z)+\overline{\langle \theta \rangle}_{\color{black}H}(1-z)=0.$$
Within the present framework, this symmetry is recovered when $A=B=C=0$. Consequently, the volume-averaged temperature vanishes, $\overline{\langle \theta \rangle}=0$, and the mean temperature at mid-height satisfies $\overline{\langle \theta \rangle}_{\color{black}H}(z=1/2)=0$.
By contrast, when $A\neq 0$, the nonlinear (quadratic) equation of state breaks this vertical symmetry.
\textcolor{black}{The latter is illustrated in figure~\ref{fig:comparison_rbc}, where we plot the difference between the horizontally-averaged temperature profiles from CDC and standard RBC, 
${\overline{\left<\theta\right>}_{\color{black}H}}_\text{CMP}
-{\overline{\left<\theta\right>}_{\color{black}H}}_\text{SRB}$. The panels show results for different Rayleigh numbers, from $\Ray=2.15\times 10^{3}$ to $\Ray=1.00\times 10^{8}$. For all conditions, we observe a loss in symmetry and a progressive net cooling as $\Ray$ increases.}

The origin of this asymmetry lies in the buoyancy term of the momentum balance (\ref{eq:momentum}), 
\begin{equation*}
b=A\theta^{2}-\theta,
\end{equation*}
which contains both linear and quadratic contributions. The quadratic contribution causes equal-and-opposite temperature anomalies to produce unequal buoyancy anomalies, thereby breaking the symmetry of the buoyancy field about the midplane $z=1/2$.

In particular, for the fixed boundary values $\theta(z=0)=-1/2$ and $\theta(z=1)=1/2$, with $0<A\leq 1$,
$$\langle b\rangle_{\color{black}H}(z=0)+\langle b\rangle_{\color{black}H}(z=1)=A/2>0 \hspace{0.25cm} \textrm{and} \hspace{0.25cm} |\langle b\rangle_{\color{black}H}(z=0)|>|\langle b\rangle_{\color{black}H}(z=1)|.$$ 
Thus, the boundary buoyancy anomalies are intrinsically asymmetric, with a larger-magnitude buoyancy anomaly near the lower boundary than near the upper boundary.
This vertical buoyancy asymmetry is, in turn, reflected in the mean temperature field.

The neutrally buoyant height, $z_n$, is defined by $$\langle b\rangle_{\color{black}H}(z=z_{n})=0, \hspace{0.25cm}\textrm{i.e.}\hspace{0.25cm}
A\langle \theta^{2}\rangle_{\color{black}H}=\langle \theta\rangle_{\color{black}H} \hspace{0.25cm}\textrm{at}\hspace{0.25cm} z=z_{n}.$$
Since $\langle \theta^{2}\rangle_{\color{black}H}>0$ is non-negative, and strictly positive in any non-trivial convective state, it follows for $A>0$ that $\langle \theta\rangle_{\color{black}H}> 0$ at $z=z_n$. The neutrally buoyant level, therefore, does not coincide with the level at which the mean temperature changes sign, and the classical midplane symmetry is lost.

To characterise this shift, we define $z_m$ as the height at which $\langle \theta\rangle_{\color{black}H}=0.$
Writing $\tilde{\theta}=\theta-\langle \theta\rangle_{\color{black}H}$, we have
$$\langle \theta^{2}\rangle_{\color{black}H}=\langle \tilde{\theta}^{2}\rangle_{\color{black}H}+\left(\langle \theta\rangle_{\color{black}H}\right)^{2},$$
{\color{black}were $\langle \tilde{\theta}^{2}\rangle_{\color{black}H}$ is the vertical profile of temperature variance.}
Therefore, at $z=z_m$, where $\langle\theta\rangle_{\color{black}H}=0$,
$$\langle b\rangle_{\color{black}H}=A\langle \tilde{\theta}^{2}\rangle_{\color{black}H}.$$
More generally, 
$$\langle b\rangle_{\color{black}H}<A\langle \tilde{\theta}^{2}\rangle_{\color{black}H} \hspace{0.25cm} \Rightarrow \hspace{0.25cm} \langle \theta\rangle_{\color{black}H}>0, \hspace{0.25cm}\textrm{whereas}\hspace{0.25cm} \langle b\rangle_{\color{black}H}>A\langle \tilde{\theta}^{2}\rangle_{\color{black}H} \hspace{0.25cm}\Rightarrow\hspace{0.25cm}\langle \theta\rangle_{\color{black}H}<0.$$
For most of the range of $\Ray$ considered here, we find that $z_m>1/2$. Hence, negative values of $\langle \theta\rangle_{\color{black}H}$ occupy more than half of the fluid depth, consistent with a negative volume-averaged temperature.
A narrow exception occurs near $\Ray \approx 4.64\times 10^{3}$, where $z_m<1/2$ and the mean temperature becomes positive.
This interval marks the transition from weak convection, with no clearly developed mixed layer, to a more vigorous regime with a well-mixed interior. 

\begin{figure}[t]
\includegraphics{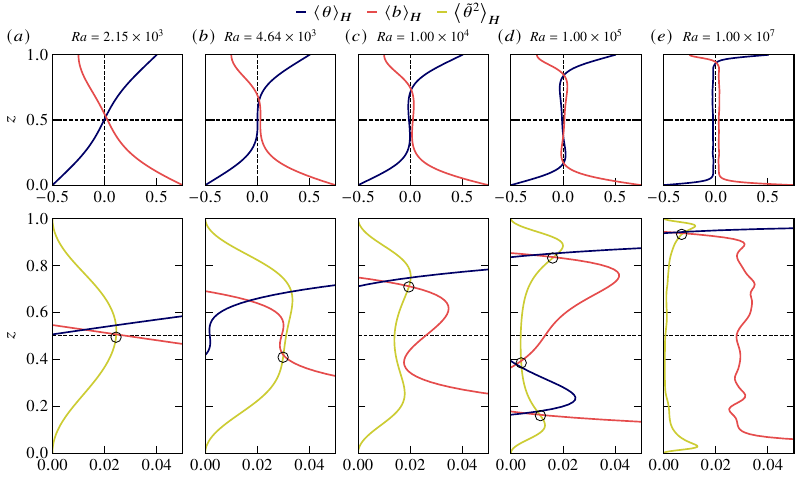}
\caption{\textcolor{black}{Instantaneous} horizontally-averaged temperature, buoyancy, and temperature variance for increasing $\Ray$, showing reduced upper-lower asymmetry and an upward shift of the zero-crossing of the mean temperature $\langle\theta\rangle_{\color{black}H}$, from positive to negative bulk mean temperature.
The intersection between $\langle b \rangle_{\color{black}H}$ and $\langle\tilde{\theta}^{2}\rangle_{\color{black}H}$ that determines the change in temperature sign is highlighted with black circles.}
\label{fig:fluctuations}
\end{figure}

The latter analysis is illustrated in \hyperref[fig:fluctuations]{figure~\ref{fig:fluctuations}}, where we display three horizontally averaged quantities, representative of the quasi-steady-state of the system, as functions of height, $z$: $\langle\theta\rangle_{\color{black}H}$, $\langle b \rangle_{\color{black}H}$, and $\langle\tilde{\theta}^{2}\rangle_{\color{black}H}$.
{\color{black} The top panels show the full extent of the profiles $\langle\theta\rangle_{\color{black}H}$ and $\langle b \rangle_{\color{black}H}$.
The bottom panels restrict the attention to positive values close to zero, emphasizing the structure of the profiles in the mixed layer.} 
Each column corresponds to a different Rayleigh number, ranging from near-supercritical conditions, $\Ray=2.15\times 10^{3}$, to turbulent conditions, $\Ray=10^{7}$.
{\color{black}As $\Ray$ increases, the horizontally averaged temperature profile $\langle\theta\rangle_{\color{black}H}$ (blue line) shows progressive symmetry breaking, highlighted in the bottom panels.}
The horizontally averaged buoyancy $\langle b\rangle_{\color{black}H}$ \textcolor{black}{(red line)} highlights significant differences between the magnitudes of the upper and lower buoyancy anomalies.
 The bottom row of \hyperref[fig:fluctuations]{figure~\ref{fig:fluctuations}} shows the intersection between $\langle b\rangle_{\color{black}H}$ and $\langle\tilde{\theta}^{2}\rangle_{\color{black}H}$, which for $A=1$ determines the depth at which $\langle\theta\rangle_{\color{black}H}$ changes sign.
This intersection shifts from the lower half of the domain in \hyperref[fig:fluctuations]{figure~\ref{fig:fluctuations}}($c$), corresponding to a positive mean temperature, to the upper half of the domain in \hyperref[fig:fluctuations]{figure~\ref{fig:fluctuations}}($c$--$e$), corresponding to a negative mean temperature $\theta_m$.
The case $\Ray = 10^5$, \hyperref[fig:fluctuations]{figure~\ref{fig:fluctuations}}($d$), \textcolor{black}{exhibits a distinct structure. Unlike the other cases, the condition $\langle\theta\rangle_{H}(z=z_{m})=0$, equivalently $\langle b\rangle_{H}=\langle\tilde{\theta}^{2}\rangle_{H}$ (see circles), is satisfied at three different depths $z_m$: two below $z=1/2$ and one above. Despite the existence of multiple zero crossings, the average temperature remains negative.}

\begin{figure}[t]
\includegraphics{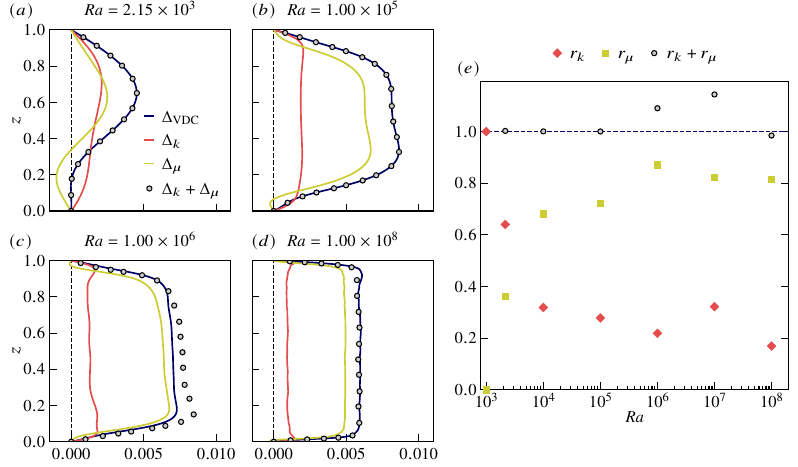}
\caption{Panels from $(a)$ to $(d)$ show the individual and joint effects of variable conductivity and viscosity in the laterally averaged temperature profile, with respect to the CDC case (\ref{eq:delta}), for different Rayleigh numbers.
Panel $(e)$ shows the relative effect of variable conductivity ($r_k$) and viscosity ($r_\mu$) in the temperature increase with respect to their joint effect (\ref{eq:ratio}), as a function of the Rayleigh number.
To emphasise the nonlinear interaction between variable viscosity and conductivity, the linear superposition of their individual effects is shown with grey circles in all panels.}
\label{fig:comparison}
\end{figure}

\subsection{Effect of temperature-dependent viscosity and conductivity on the temperature field}
The results summarised in \hyperref[fig:mean_theta]{figure~\ref{fig:mean_theta}} show that variable conductivity ($k$) and viscosity ($\mu$) raise the mean temperature of the system with respect to the CDC case.
However, they do not allow for weighing their relative contributions in this increase.
An advantage of our framework is that we can explore these effects independently, as the condition $B = 0$ deactivates temperature dependence in viscosity, while $C = 0$ does the same for conductivity -- see the right side of equations (\ref{eq:momentum}) and (\ref{eq:heat}).
Then we can compare the different scenarios using the CDC case, that is $B=C=0$, as a reference.

In particular, we consider the difference between the time-averaged temperature profiles
\begin{equation}
\Delta_i\left(z\right) = {\overline{\left<\theta\right>}_{\color{black}H}}_i
-{\overline{\left<\theta\right>}_{\color{black}H}}_\text{CDC}
\label{eq:delta}
\end{equation}
for $i=k$ corresponding to the case $B=0$, $i=\mu$ to $C=0$ and $i=\text{VDC}$ for $B$ and $C$ different from zero.
Moreover, we can compare $\Delta_\text{VDC}$ with the sum $\Delta_k + \Delta_\mu$ to investigate the extent to which the joint effect of $k$ and $\mu$ can be explained from a linear superposition of individual contributions.
The aforementioned depth-dependent comparison is presented in \hyperref[fig:comparison]{figure~\ref{fig:comparison}}, increasing the Rayleigh number from panels $(a)$ to $(d)$. \textcolor{black}{The vertical axes denote $z$ while $\Delta_i$ is shown in the horizontal axes.}
To {\color{black}measure} the relative contributions of $k$ and $\mu$ {\color{black}with respect to the} global temperature increase, we define
\begin{equation}
r_i = \frac{\overline{\left<\theta\right>}_i
-{\overline{\left<\theta\right>}}_\text{CDC} }
{\overline{\left<\theta\right>}_\text{VDC}
-{\overline{\left<\theta\right>}}_\text{CDC}},
\label{eq:ratio}
\end{equation}
shown in \hyperref[fig:comparison]{figure~\ref{fig:comparison}}($e$) as a function of $\Ray$. As in the previous case, we also consider the linear superposition $r_k + r_\mu$.

We first note that variable conductivity consistently produces a local increase in temperature along $z$.
This is mainly explained by the positive term $\nabla\theta \bcdot \nabla \theta$ on the right side of (\ref{eq:heat}), which can be interpreted as an internal heat source.
In contrast, the effect of viscosity can produce local decreases in temperature, despite the fact that the net result is still an increase in the mean temperature of the system.
For example, cases in the range $\Ray \leq 10^5$ show a local temperature decrease in the bottom boundary layer -- see panels ($a$) and ($b$) of \hyperref[fig:comparison]{figure~\ref{fig:comparison}}.
This drop is explained by the viscosity increase near the bottom, which ultimately enhances conductive heat transport from the cold boundary.
This effect disappears when $\Ray$ increases, as the boundary layers become thinner and more turbulent.
However, the dominating mechanism differs in developed convective layers: \textcolor{black}{warm downwelling plumes generated at the top boundary form in a lower-viscosity region than cold upwelling plumes generated at the bottom boundary}.
Consequently, the convective heat transport from the cold bottom is weakened, while the one from the warmer top is enhanced, resulting in a net temperature increase in the well-mixed region.

Second, we note that, for cases close enough to the onset of convection, the effect of conductivity dominates the temperature increase.
This is consistent with the fact that the conductive state is independent of viscosity and controlled by conductivity.
However, as the Rayleigh number increases, the temperature shift is dominated by the effect of variable viscosity. This result is expected from the values of the coefficients $B$ and $C$, the former being two orders of magnitude bigger than the latter.

Finally, we discovered that for $\Ray\leq 10^5$ -- the range of steady or weakly transient convection -- the joint effect of viscosity and conductivity in the temperature profile coincides with the linear superposition of their individual effects.
This equivalence is notable, considering that variable \textcolor{black}{diffusion coefficients} introduce nonlinear terms in the heat (\ref{eq:heat}) and momentum (\ref{eq:momentum}) balances.
For higher Rayleigh numbers, the nonlinearities modify -- either increasing or decreasing -- the temperature shift.
These nonlinear effects are relatively small for \textcolor{black}{$\Ray=10^8$}, while they contribute significantly to the increase in the mean temperature in the transition regime, for Rayleigh numbers \textcolor{black}{between $10^6$ and $10^7$}.

\subsection{Scaling relations}
For classical OB RBC with rigid isothermal boundaries -- the case $A=B=C=0$ in our framework -- linear stability theory gives $Ra_c=1707.76$ \citep[e.g.,][]{chandrasekhar2013hydrodynamic}.
In the CDC cases, however, the critical Rayleigh number is reduced.
This is consistent with the nonlinear equation of state, which renders the destabilising buoyancy production non-uniform across the layer.
In particular, the strongest unstable stratification is concentrated near the lower boundary, controlling the onset.
To show this, we define a {\color{black}depth-dependent} Rayleigh number based on the density gradient {\color{black}computed from the conductive temperature profile $\varphi(z)$ (see equation \ref{eq:cond_profile}):}
\begin{equation}
\Ray_\ell^\text{CDC} (z) = \Ray_\ell^\text{CDC} (\varphi) = \frac{gh^3\rho_rc_v}{\mu_r k_r}
\frac{\mathrm{d}\rho}{\mathrm{d}z} = \left(1 - 2A\varphi\right)\Ray .
\end{equation}
We first note that $\Ray_\ell^\text{CDC}(\varphi=0) = \Ray$, that is, in the conductive state, $\Ray$ captures the local density gradient at the reference temperature.
Second, we note that
\begin{equation*}
\Ray_\ell^\text{CDC}(\varphi=-1/2) / \Ray = 1 + A \geq 1.
\end{equation*}
That is, unstable density gradients are higher at the bottom, an asymmetry that is controlled by the parameter $A$.
To account for the more unstable bottom, $\Ray_c$ has to be smaller when $A > 0$.
For the VDC cases, the viscosity and conductivity are depth-dependent, having opposite effects. Viscosity attains its maximum at the bottom boundary, stabilising it, while conductivity is minimum at such a boundary, rendering it less stable.
To weigh the relevance of these mechanisms, we generalise the previous definition of the local Rayleigh number considering temperature-dependent viscosity (\ref{eq:viscosity}) and conductivity (\ref{eq:conductivity}),
\begin{equation}
\Ray_\ell^\text{VDC} (\varphi) = \frac{gh^3\rho_rc_v}{\mu k}
\frac{\mathrm{d}\rho}{\mathrm{d}z} = \frac{\left(1 - 2A\varphi\right)}{\left(1-B\varphi\right)\left(1+C\Pran\varphi\right)^2}\, \Ray,
\end{equation}
and compute
\begin{equation*}
\frac{\Ray_\ell^\text{VDC}(\varphi=-1/2)}{\Ray}  \approx \frac{1 + A}{\left(1.07\right)\left(0.98\right)} \approx 0.95 \, \frac{\Ray_\ell^\text{CDC}(\varphi=-1/2)}{\Ray}.
\end{equation*}
From the point of view of the onset of convection in our parameter range, the stabilising effect of viscosity at the bottom is more important than the opposite effect of conductivity, increasing $\Ray_c$ with respect to the CDC case.
In fact, for the VDC case, we obtain $\Ray_c \approx 1707$, very close to the critical Rayleigh number of {\color{black}standard} RBC.

The different critical Rayleigh numbers explain why we have used a supercriticality $\epsilon = \Ray-\Ray_c$ to make scaling laws comparable between the CDC and VDC cases.
In particular, Figure~\ref{fig:nu}($a$) shows the heat-transfer scaling, where two distinctive regions are apparent.
Close to onset, weakly nonlinear theory gives $\xi\sim \epsilon^{1/2}$ for the amplitude $(\xi)$ of the primary convective mode \citep{cross1980derivation,cross1993pattern}.
Since the convective heat flux is quadratic in $\xi$, the excess Nusselt number scales as $$\Nuss-1 \propto \xi^2 \propto \epsilon \propto (\Ray-\Ray_c).$$
Our low-supercriticality data $(\epsilon\lesssim 10^2)$, showing the exponent close to unity, are consistent with this prediction.
At larger forcing $(\epsilon\gtrsim 10^5)$, the data are well described by the effective law $$\Nuss-1 \propto (\Ray-\Ray_c)^{\alpha},$$ with $\alpha$ lying within the range commonly reported for buoyancy-driven convection, namely between the standard $2/7$ and $1/3$ scalings \citep{grossmann2000scaling}.

These two regions are also observed in the scaling for the kinetic energy -- here described by the Reynolds number ($\Ray$) -- as shown in Figure~\ref{fig:nu}($b$).
Near the onset, the characteristic velocity is proportional to the mode amplitude, which gives $$\Rey \propto \xi \propto \epsilon^{1/2}, \qquad\text{or equivalently}\qquad \Rey^2 \propto \Ray-\Ray_c.$$
Thus, the low-$\Ray$ branch is consistent with a viscously controlled, weakly nonlinear regime.
{\color{black}As mention earlier, at a larger Rayleigh number, the $Re$--$Ra$ relation steepens slightly, following a scaling previously observed by \cite{van2013comparison} for 2D convection}.

\section{Concluding remarks and implications}\label{sec:conclusion}

\textcolor{black}{Here, we show that Rayleigh-Bénard convection (RBC) in coldwater, between the freezing point $T_{\textit{fp}}=0^{\circ}\textrm{C}$ and the temperature of maximum density $T_{\textit{md}}\approx3.98^{\circ}\textrm{C}$, breaks local and global symmetries associated with the thermal field that characterizes the standard RBC model.} These asymmetries arise from the combined action of the nonlinear equation of state (EOS) characterising the anomalous temperature-density relationship near the freezing point and temperature-dependent thermal conductivity and viscosity.

In this coldwater regime, the nonlinear EOS reduces the buoyancy generated by a given temperature fluctuation as the fluid approaches $T_\textit{md}$, thereby making the flow especially sensitive to departures from constant material properties. Coldwater convection should therefore not, in general, be regarded as {\color{black} a ``classic Oberbeck-Boussinesq fluid'' in the strictest sense: incompressible, with a linear equation of state and constant diffusion coefficients and heat capacity.} 

Returning to the questions posed at the outset, three main conclusions emerge from this study. First, the quadratic EOS fundamentally affects the buoyancy symmetry in the system. Moreover, even the modest temperature dependence of viscosity and thermal conductivity present in coldwater between the freezing point $T_\textit{fp}$ and $T_\textit{md}$ affects both the conductive base state and the consequent convective state. \textcolor{black}{These effects produce measurable changes in the thermo-fluid dynamics, including shifts in the mean temperature $\theta_m$, up to $0.08^{\circ}\textrm{C}$ for Rayleigh numbers ($\Ray$) of order $10^{8}$
}, which is associated with a loss of the vertical symmetry in the mean temperature profile $\langle \theta \rangle_{\color{black}H}$. While this temperature shift plateaus in the studied range of $\Ray$, future studies could explore how this temperature shift evolves for higher $\Ray$. {\color{black}
{\color{black}For high-$\Ray$ we found that the temperature shift is primarily controlled by the nonlinear EOS, with a temperature-dependent diffusion coefficient providing secondary corrections. The latter supports the use of models with nonlinear EOS but constant diffusion coefficients for coldwater bodies.}
Furthermore, our analysis shows that when considering only the effect of the nonlinear EOS, with constant diffusion coefficients, the critical Rayleigh number ($\Ray_c$) for the onset of convection decreases. However, when including the effects of temperature-dependent diffusion coefficients, the estimated $\Ray_c$ returns to approximately the standard value of $1708$.}

Second, we found that temperature-dependent thermal conductivity exerts a stronger influence at low Rayleigh numbers, including in the conductive regime, where it alters the mean temperature. By contrast, temperature-dependent viscosity and the quadratic EOS become more important at higher Rayleigh numbers, where convective {\color{black} transport dominates}.

Third, despite the symmetry breaking in the temperature field, global laws for heat and momentum transport are not modified. The heat flux remains broadly consistent with classical supercritical behaviour, with $\Nuss - 1 \propto (\Ray-\Ray_{c})$ in the low-$\Ray$ regime and $\Nuss \propto (\Ray-\Ray_{c})^{0.3}$ in the high-$\Ray$ regime \citep{xia2002heat}. In addition, for coldwater at an intermediate Prandtl number, $\Pran \approx 12.626$, the $\Ray$--$\Rey$ relation follows the classic power law $\Rey \propto (\Ray-\Ray_{c})^{1/2}$ \citep{grossmann2000scaling,grossmann2001thermal}.
{\color{black}While our data shows a slightly increased exponent for $\Ray > 10^4$, this is know to occur in two-dimensional RBC models \citep{van2013comparison}}.

Our findings are directly relevant to the thermo-fluid dynamics of criosopheric waters, including fully and partially ice-covered lakes \citep{allum2024simulations,estay2026under,noto_PRF_2026}, proglacial lakes, supraglacial ponds \citep{yang2025bistability}, meltwater channel flows \citep{fernandez2021laboratory},  permafrost in porous media \citep{holzbecher1997numerical,magnani2024convection}, and more fundamental studies on growth and melting dynamics of ice bodies in contact with liquid water \citep{keitzl2016impact,couston2021topography,wang2021growth,bellincioni2025melting,noto_SAD_2026,berhanu2026self,guo2025effects}. These systems commonly operate with small temperature differences and weak buoyancy forcing, yet still sustain convection that governs the transport of heat, dissolved constituents and suspended matter \citep[e.g.,][]{yang2017high,bouffard2019under,perga2020rotiferan}. In such flows, the dynamics depend sensitively on how efficiently small thermal anomalies are converted into buoyancy near the temperature of maximum density, so the effects of the nonlinear EOS remain dynamically important even when the variations in {\color{black}other material properties} are modest. Beyond the effects from temperature-dependent material properties considered here, additional processes may become important in more extreme cryospheric environments; in particular, the high-pressure conditions expected in subglacial lakes can further modify coldwater convection \citep{wuest2000priori,wells2008circulation,couston2021turbulent}. Similar considerations arise in coldwater engineered systems, including food processes \citep{leung2007fluid,jiang2025advancements} and ice-covered reservoirs used for heat storage \citep{fink2014large,eggimann2023potential,yang2024enhanced}, where determining the mean liquid water temperature is relevant for food quality and available thermal energy.

The coldwater Rayleigh--Bénard system studied here, therefore, serves as a minimal framework for identifying which ingredients must be retained when modeling cold freshwater dynamics across cryospheric and engineered settings.\\

\noindent\textbf{Acknowledgments.} We thank the computational support from the SAS General Purpose Cluster at the University of Pennsylvania. \textcolor{black}{This manuscript also benefited from the constructive feedback of anonymous referees, whom we thank}.\\

\noindent\textbf{Funding.} The authors gratefully acknowledge support from the University of Pennsylvania start-up grant to H.N.U. G.E. acknowledges the SAS Dissertation Completion Fellowship, University of Pennsylvania. D.N. acknowledges the financial support from JST EXPERT-J, Japan Grant Number JPMJEX2511.\\

\noindent\textbf{Declaration of interests.} The authors report no conflicts of interest.\\

\noindent\textbf{Author contributions.} G.E., D.N. and H.N.U. designed research; G.E. performed research; G.E. derived theory; G.E. analyzed data; H.N.U. supervised research; H.N.U. acquired funding; G.E., D.N. and H.N.U. wrote the manuscript.\\

\noindent\textbf{Data availability statement.} Dedalus, the open-source, Python-based, MPI-parallel spectral solver used in this work, is available at \href{https://github.com/DedalusProject/dedalus}{https://github.com/DedalusProject/dedalus}.

\clearpage
\begin{appen}
\section{}
\begin{table}
\centering
\begin{tabular}{crrccccc||crrccccc}
$\Ray$ & $n_x$ & $n_z$ & VDC & CDC & SRB & $k$ & $\mu$ & $\Ray$ & $n_x$ & $n_z$ & VDC & CDC & SRB & $k$ & $\mu$\\[3pt]
$1690$ & 256 & 128 & - & $\checkmark$ & - & - & - & $1.00\times 10^5$ & 256 & 128 & $\checkmark$ & $\checkmark$ & $\checkmark$ & $\checkmark$ & $\checkmark$\\
$1700$ & 256 & 128 & $\checkmark$ & $\checkmark$ & $\checkmark$ & - & - & $2.15\times 10^5$ & 256 & 128 & $\checkmark$ & $\checkmark$ & $\checkmark$ & - & - \\
$1710$ & 256 & 128 & $\checkmark$ & $\checkmark$ & $\checkmark$ & - & - & $4.64\times 10^5$ & 256 & 128 & $\checkmark$ & $\checkmark$ & $\checkmark$ & - & - \\
$1714$ & 256 & 128 & $\checkmark$ & $\checkmark$ & $\checkmark$ & - & - & $1.00\times 10^6$ & 256 & 128 & $\checkmark$ & $\checkmark$ & $\checkmark$ & $\checkmark$ & $\checkmark$\\
$1722$ & 256 & 128 & $\checkmark$ & $\checkmark$ & $\checkmark$ & - & - & $2.15\times 10^6$ & 512 & 128 & $\checkmark$ & $\checkmark$ & $\checkmark$ & - & - \\
$1740$ & 256 & 128 & $\checkmark$ & $\checkmark$ & - & - & - & $4.64\times 10^6$ & 512 & 128 & $\checkmark$ & $\checkmark$ & $\checkmark$ & - & - \\
$1800$ & 256 & 128 & $\checkmark$ & $\checkmark$ & $\checkmark$ & - & - & $1.00\times 10^7$ & 512 & 128 & $\checkmark$ & $\checkmark$ & $\checkmark$ & $\checkmark$ & $\checkmark$\\
$2.15\times 10^3$ & 256 & 128 & $\checkmark$ & $\checkmark$ & $\checkmark$ & $\checkmark$ & $\checkmark$ & $2.15\times 10^7$ & 1024 & 256 & $\checkmark$ & $\checkmark$ & $\checkmark$ & - & - \\
$4.64\times 10^3$ & 256 & 128 & $\checkmark$ & $\checkmark$ & $\checkmark$ & - & - & $4.64\times 10^7$ & 1024 & 256 & $\checkmark$ & $\checkmark$ & $\checkmark$ & - & -  \\
$1.00\times 10^4$ & 256 & 128 & $\checkmark$ & $\checkmark$ & $\checkmark$ & $\checkmark$ & $\checkmark$ & $1.00\times 10^8$ & 1024 & 256 & $\checkmark$ & $\checkmark$ & $\checkmark$ & $\checkmark$ & $\checkmark$\\
$2.15\times 10^4$ & 256 & 128 & $\checkmark$ & $\checkmark$ & $\checkmark$ & - & - & $2.15\times 10^8$ & 2048 & 512 & $\checkmark$ & $\checkmark$ & - & - & -\\
$4.64\times 10^4$ & 256 & 128 & $\checkmark$ & $\checkmark$ & $\checkmark$ & - & - 
\end{tabular}
\caption{Summary of numerical simulations, indicating the Rayleigh number ($\Ray$), horizontal nodes ($n_x$) and vertical ($n_z$) nodes for the following cases: variable diffusion coefficients (VDC), constant diffusion coefficients (CDC), standard Rayleigh--Bénard (SRB), nonlinear EOS with constant viscosity and variable conductivity ($k$), nonlinear EOS with constant conductivity and variable viscosity ($\mu$). A tick ($\checkmark$) indicates that a case was run for a particular $\Ray$, while a hyphen (-) indicates that it was not.}
\label{tab:simulations}
\end{table}

\begin{figure}[ht]
\includegraphics[]{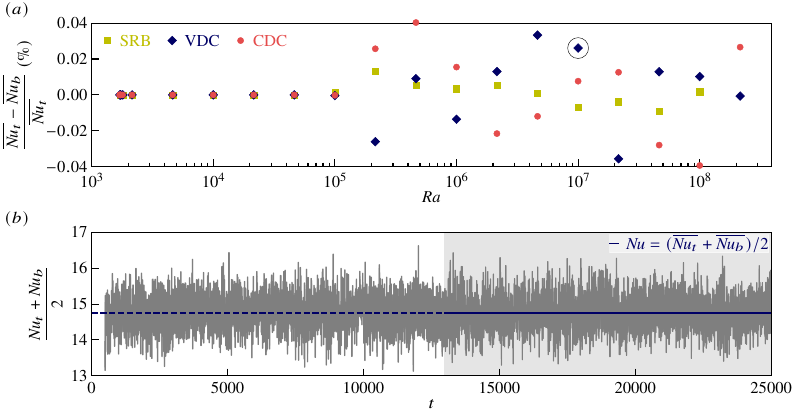}
\caption{$(a)$ Validation of steady-state conditions for SRB (yellow squares), VDC (blue diamonds) and CDC (red circles) cases.
$(b)$ Computation of Nusselt number $\Nuss$ (horizontal blue line) from a time series (grey line) over the shaded time window, for the case marked with a grey ring on panel $(a)$. See equation (\ref{eq:nuss_alt}).}
\label{fig:steady}
\end{figure}

\end{appen}
\clearpage

\bibliographystyle{jfm}
\bibliography{jfm-ref}

\end{document}